\documentclass[twocolumn,
amsmath,
amssymb,
aps,
nofootinbib,
raggedbottom,
superscriptaddress,
nobalancelastpage,prl]{revtex4-2}
\usepackage{graphicx} 
\usepackage{preamble}
\usepackage{color}
\usepackage[dvipsnames]{xcolor}
\usepackage{cancel}

\makeatletter
\@booleantrue\acknowledgments@sw
\makeatother

\providecommand{\CC}{\mathbb{C}}
\providecommand{\RR}{\mathbb{R}}

\providecommand{\SU}{\mathrm{SU}}

\providecommand{\tr}{\mathrm{tr}}
\newcommand{\CP}{\mathbb{C}P^{N-1}}
\newcommand{\calM}{\mathcal{M}}
\newcommand{\calE}{\mathcal{E}}

\newcommand{\dd}{\mathrm{d}}
\newcommand{\ii}{\mathrm{i}}

\newcommand{\eps}{\epsilon}
\newcommand{\hyp}{{}_0F_{N-1}}
\newcommand{\bpsi}{\bar\psi}
\newcommand{\sdd}{\mathrm{sd}}

\begin{document}

\title{Fractional $\theta$-dependence from the sum over all instanton sectors}

\author{Yui Hayashi}
\affiliation{Department of Physics, The University of Tokyo, Hongo, Bunkyo-ku, Tokyo 113-0033, Japan}
\author{Mithat \"Unsal}
\affiliation{Department of Physics, North Carolina State University, Raleigh, NC 27607, USA}

\date{\today}

\begin{abstract}

An instanton carries integer topological charge, seemingly at odds with the multibranched $\theta$ dependence of the Yang--Mills vacuum. We show that summing over all instanton sectors resolves this mismatch in self-dual supersymmetric $\mathbb{C}P^{N-1}$ and $\mathrm{SU}(N)$ Yang--Mills theories with a chiral mass, in two and four dimensions.
The path integral reduces to a sum of instanton moduli-space volumes, yielding the multibranched vacuum structure.
In four dimensions, this extends Nekrasov's instanton counting for Seiberg--Witten theory to a QCD-like theory without scalars.

\end{abstract}

\maketitle

\section{Introduction}
Instantons \cite{BPST,tHooft76,tHooft76b} were discovered with the hope that they would explain the vacuum of asymptotically free theories: the $\theta$-dependence, chiral symmetry breaking, confinement and the mass gap \cite{CDG,Polyakov75,Polyakov77}. That hope was abandoned within a few years, and it is  accepted  that whatever confines quarks, it is not instantons \cite{ColemanBook}. Two reasons were made sharp in \cite{Witten79,Affleck80}. First, the one-instanton integral is infrared divergent: an instanton of size $\rho$ carries a measure $\rho^{b_0-d-1}\dd\rho$, with $b_0$ the one-loop coefficient, so large instantons dominate and the dilute gas is not under control. Second, an instanton weighs $e^{-S_I}$ with $S_I=NS_0\propto N/\lambda$ at fixed 't~Hooft coupling, so instanton effects appear to be $O(e^{-N})$, while the large-$N$ solution of the $\CP$ model \cite{DLD} and the expected structure of Yang--Mills theory \cite{WittenTheta} give a vacuum energy $\min_\ell f\big((\theta+2\pi\ell)/N\big)$, with $N$ branches. 
Since then \cite{ColemanBook,Witten79, Affleck80}, the conclusion that became standard is that semiclassics has no standing as a first-principles account of the vacuum on $\mathbb R^d$.

However, reliable semiclassical studies on $\RR^3\times S^1$ \cite{Unsal2008,Unsal2009,ShifmanUnsal2008,UnsalYaffe2008,Unsal2020,Poppitz2022} and $\RR^2\times T^2$ \cite{TanizakiUnsal2022,HayashiTanizaki2024a,HayashiTanizaki2024b}, and on $\RR\times S^1$ in two dimensions \cite{DunneUnsal2013,Bruckmann,Brendel,MisumiNittaSakai2014,Fujimori2019}, with judiciously chosen boundary conditions, where the  instantons fractionate into $N$ constituents \cite{LeeYi97,LeeLu98,KraanvanBaal98a,KraanvanBaal98b}, showed that  these fractional instantons and their composites (bions) produce the $\theta$-dependence, the mass gap and confinement. Besides providing solutions of these theories in weakly coupled domains that are adiabatically connected to the strongly coupled theory on $\RR^d$ \cite{ACD2021,BCD2018, BPU2018}, this raises a deeper issue:  that the textbook picture of instantons on $\RR^d$ \cite{ColemanBook} may actually be too naive.

We wish to understand the role of instantons on $\RR^d$ more properly, and the lack of success in doing so leads us to new tools. It has recently been realized that the self-dual (Chalmers--Siegel) theory  \cite{Parkes92,ChalmersSiegel}, the path integral in which the anti-self-dual part of the field strength is set to zero by a Lagrange multiplier, 
may provide new non-perturbative tools \cite{HU,DGNU}. There is a vast literature on the perturbative uses of self-dual Yang--Mills theory, which is known mainly as a one-loop-exact perturbative device, behind the all-plus amplitudes, the maximally-helicity-violating expansion, the double copy and celestial chiral algebras \cite{Bardeen96,Cangemi96,BCDK94,Mahlon94,WittenTwistor,CSW04,Mansfield06,Mason05,BMS07,MonteiroOConnell11,Costello21,CostelloPaquette22}. Its utility as a non-perturbative tool is new, and it is introduced here more systematically. The self-dual theory keeps the instantons, the one-loop-exact running coupling and the strong scale $\Lambda$, and it can be continuously  deformed to the physical theory \cite{HU,DGNU, HRSS2026}; it drops the anti-instantons and all perturbative corrections beyond one loop \cite{Frenkel:2006fy,Frenkel:2008vz}. It is to the physical theory what the holomorphic data is to a supersymmetric one.

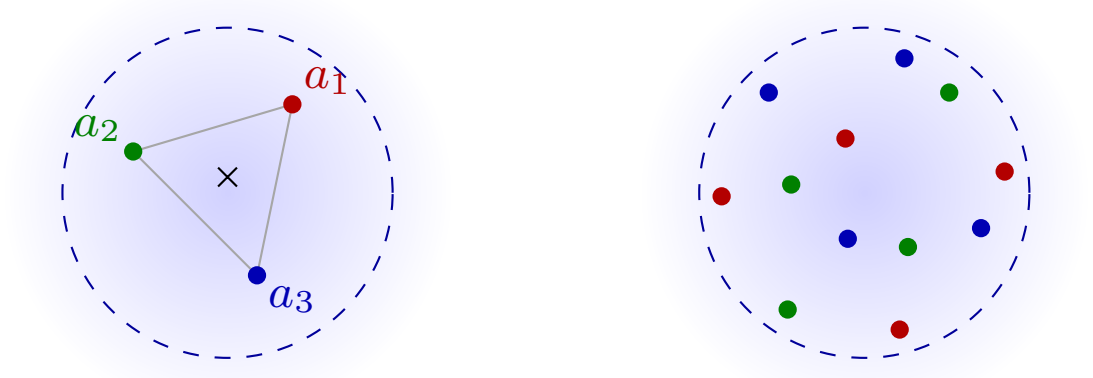
\begin{figure}[t]
\centering
\begin{tikzpicture}[scale=0.78]
\begin{scope}
\shade[inner color=blue!18, outer color=white] (0,0) circle (1.9);
\draw[blue!60!black, dashed] (0,0) circle (1.4);
\coordinate (a1) at (0.55,0.75);
\coordinate (a2) at (-0.8,0.35);
\coordinate (a3) at (0.25,-0.7);
\draw[gray!70] (a1)--(a2)--(a3)--cycle;
\fill[red!70!black]   (a1) circle (2.2pt);
\fill[green!50!black] (a2) circle (2.2pt);
\fill[blue!70!black]  (a3) circle (2.2pt);
\node[font=\small,red!70!black]   at ($(a1)+(0.3,0.2)$)  {$a_1$};
\node[font=\small,green!50!black] at ($(a2)+(-0.3,0.2)$) {$a_2$};
\node[font=\small,blue!70!black]  at ($(a3)+(0.3,-0.2)$) {$a_3$};
\coordinate (ac) at ($1/3*(a1)+1/3*(a2)+1/3*(a3)$);
\draw[black] (ac) ++(-0.08,-0.08)--++(0.16,0.16) ++(-0.16,0)--++(0.16,-0.16);
\node[font=\small] at (0,-2.3) {one instanton, $N=3$};
\end{scope}
\begin{scope}[xshift=5.4cm]
\shade[inner color=blue!18, outer color=white] (0,0) circle (1.9);
\draw[blue!60!black, dashed] (0,0) circle (1.4);
\foreach \x/\y in {-0.16/0.46,0.30/-1.16,1.19/0.18,-1.21/-0.03}{\fill[red!70!black] (\x,\y) circle (2.2pt);}
\foreach \x/\y in {0.72/0.85,-0.62/0.07,-0.65/-0.99,0.37/-0.46}{\fill[green!50!black] (\x,\y) circle (2.2pt);}
\foreach \x/\y in {-0.14/-0.39,0.34/1.14,-0.81/0.85,0.99/-0.30}{\fill[blue!70!black] (\x,\y) circle (2.2pt);}
\node[font=\small] at (0,-2.3) {$Q=4$ instantons, $N=3$};
\end{scope}
\end{tikzpicture}
\caption{  Left: a $\CP$ instanton is a single lump of topological charge (the cross is its center), but its moduli are $N$ points $a_i$ in the plane: for $\phi=[z-a_1:\cdots:z-a_N]$, the $i$-th component of $\phi$ vanishes at $a_i$ \cite{NU}; a trap (shaded) effectively acts on each point independently.
 Right: $Q$ instantons are $N$ species (colors) of $Q$ points, the roots of the $N$ polynomials of the holomorphic map. When the lumps overlap, a point cannot be assigned to a particular lump, and the gas has the entropy of $N$ species, each with $Q$ identical particles.}
\label{fig:picture}
\end{figure}

In this Letter we perform the sum over all instanton sectors in two self-dual theories with a chiral mass: the $\mathcal N=(2,2)$ $\CP$ model in two dimensions, where the sum closes in a classical special function for every $N$, and $\mathcal N=1$ $\SU(N)$ super-Yang--Mills (SYM) theory in four dimensions, where we evaluate it in a large-volume limit of Nekrasov--Shatashvili (NS) type. 
It was argued that instanton descriptions are incompatible with the large-$N$ expansion \cite{Witten79}; however, as we show below, a sum over topological sectors, which can be done exactly without relying on the dilute gas approximation, can reproduce the desired fractional $\theta$ dependence.
In both theories the free energy of the full sum is extensive, $\log Z\propto V\,u^{1/N}$, and the vacuum is multibranched in $\theta$.
The mechanism is statistical. In two dimensions the moduli of $Q$ instantons are $N$ species of $Q$ points in the plane \cite{NU}, and when instantons overlap the sum becomes a gas of $N$ species, each with $Q$ identical particles (Fig.~\ref{fig:picture}). Each particle has weight $u^{1/N}\propto e^{-S_I/N+\ii\theta/N}$, charge $1/N$ and two fermion zero modes: the quantum numbers of a fractional instanton, without any configuration of fractional action.

\section{Two dimensions: self-dual $\mathcal N=(2,2)$ $\CP$ with a chiral mass}

As a simple illustration, we consider self-dual $\mathcal N=(2,2)$ $\CP$ model with a chiral mass.
A main motivation for studying this model is that its partition function involves the sum over all instanton sectors like the bosonic self-dual theory, but each sector reduces to a simple moduli space integral.

We use the first-order formalism of \cite{HU}. The field $\phi\in\CC^N$, $|\phi|^2=1$, modulo phases, has the composite connection $A_\mu=-\ii\phi^\dagger\partial_\mu\phi$ and topological charge $Q=\frac1{2\pi}\int F$; with $D_\pm=D_1\pm\ii D_2$, the self-duality equation $D_+\phi=0$ says that $\phi$ is a holomorphic map $\CC\to\CP$. The $\mathcal{N}=(2,2)$ fermions $\psi_\pm,\bpsi_\pm$ live in the tangent bundle; in a charge-$Q$ background $\psi_-$ and $\bpsi_+$ carry $NQ$ zero modes each. The self-dual theory keeps the terms linear in the multipliers $(h,\bpsi_-,\psi_+)$ and adds a chiral mass for the two zero-mode fields:
\begin{align}
 S=\int\!\dd^2x\Big[&-\ii h^\dagger D_+\phi-\ii(D_+\phi)^\dagger h-\ii\bpsi_-D_+\psi_-\nonumber\\
 &+\ii\bpsi_+D_-\psi_++m\,\bpsi_+\psi_-\Big]+\int\!\frac{\dd^2x}{2\pi}\,T(x)\,F,
 \label{eq:S2d}
\end{align}
with $T(x)=2\pi/g_\sdd^2-\ii\theta+\eps_1|z|^2$.

Three remarks are in order. (i) One can extrapolate to the full $\mathcal{N}=(2,2)$ theory by adding the terms quadratic in the multipliers, $\epsilon^2|h|^2$ and the four-fermion terms, and dialing $\epsilon\to g$. This is the $\epsilon$-deformation of \cite{HU}; without it only instantons, $Q\ge0$, survive. (ii) $m$ is a chiral mass in the sense of a source for $\bpsi_+\psi_-$. With both $m$ and the anti-chiral mass $m_2$ of $\bpsi_-\psi_+$, the fermion pole is at $p^2=-m\,m_2$, so the physical mass is $\sqrt{m\,m_2}$; in the chiral-mass limit $m_2\to0$ the fermions become massless and $m$ probes the chiral condensate. 
Note that this system has the remaining half supersymmetry.
The chiral mass with the trap potential $m \bar{\psi}_+ \psi_- + \epsilon_1 |z|^2 F /2\pi$ is invariant under a modified supersymmetry $\mathcal{Q}_{\epsilon}$, where $\mathcal{Q}_{\epsilon}^2$ generates the rotation, which can be used for the localization technique.
(iii) The last term of $T$ is an infrared regulator, a position-dependent topological coupling that suppresses instantons far from the origin by $e^{-\eps_1|z|^2}$.\footnote{\label{footnote:IRreg} 
Each $Z_Q$ in \eqref{eq:Zmod2d} is the equivariant volume of $\calM_Q$ under rotations, and it is finite for $\eps_1>0$. On $\RR^2$ the size modes are not normalizable \cite{Ward,Speight}: along them the $L^2$ metric and the trap $H$ both grow like the logarithm of the cut-off. The two logarithms cancel in the equivariant volume, which depends only on the rotation weights at the fixed point. With a $\mathcal Q_\epsilon$-invariant cut-off at $|z|\sim\delta^{-1/2}$, the result is \eqref{eq:ZF} up to terms suppressed by $e^{-\eps_1/\delta}$. See
the longer version \cite{SDSYM} for details.
}
Since $\int T\,F/2\pi$ is a function on the moduli space, it never enters a fluctuation operator.
The parameter $\eps_1$ plays the role of the Nekrasov--Shatashvili $\hbar$.

The coupling $2\pi/g_\sdd^2$ runs at one loop with coefficient $N$ \cite{HU}, $\Lambda^N=\mu^Ne^{-2\pi/g_\sdd^2(\mu)}$. A charge-$Q$ configuration has $2NQ$ bosonic and $2NQ$ fermionic zero modes, which in Pauli--Villars regularization contribute $\mu^{2NQ}\mu^{-NQ}$; with $e^{-2\pi Q/g_\sdd^2+\ii Q\theta}$ this is $\Lambda^{NQ}e^{\ii Q\theta}$, and the chiral mass contributes $m^{NQ}$. Hence the instanton weight is
\begin{equation}
u=(C_N\,m\Lambda)^N e^{\ii\theta},\qquad [u]=2N,
 \label{eq:u2d}
\end{equation}
with $C_N$ a per-instanton constant. Only $\theta+N\arg m$ is physical, and we take $m>0$.

The free energy density has dimension two, so if $\log Z$ is extensive and analytic in $u$, then $\log Z/V\propto u^{1/N}=C_Nm\Lambda\,e^{\ii\theta/N}$ for a two-dimensional volume $V$. What follows shows that extensivity holds, and how.

The multipliers appear linearly in \eqref{eq:S2d}. Integrating $h$ gives $\delta(D_+\phi)$, so $\int D\phi$ localizes on the moduli space $\calM_Q$ of degree-$Q$ holomorphic maps, with Jacobian $1/|\det{}'D_+|^2$; integrating $\bpsi_-$ gives $\delta(D_+\psi_-)$ and $\det{}'D_+$, and $(\psi_+,\bpsi_+)$ give $\det{}'D_-=\overline{\det{}'D_+}$. The determinants cancel: no Gaussian integral is left, and there are no higher loops. The mass sees only the zero modes, on which it is the K\"ahler form of $\calM_Q$, so the Grassmann integral is exactly $m^{NQ}$. The topological term splits into the action of charge $Q$, absorbed in $u^Q$, and a function on the moduli space:
\begin{equation}
Z=\sum_{Q\ge0}u^Q\!\int_{\calM_Q}\!\dd{\rm vol}_{L^2}\,e^{-H},\quad H=\eps_1\!\int\!\dd^2x\,|z|^2q,
 \label{eq:Zmod2d}
\end{equation}
with $q=F/2\pi$ the topological density. Each sector is the volume of the moduli space of charge-$Q$ maps, weighted by the trap.

\emph{The sum over all topological sectors.} 
Fix the vacuum at infinity, $\phi(\infty)=[1:\cdots:1]$ \cite{NU}. A degree-$Q$ map is then $\phi=[P_1:\cdots:P_N]$ with every
\begin{equation*}
 P_a(z)=\prod_{\alpha=1}^Q(z-a_{a,\alpha}),\qquad a=1,\dots,N,\quad \alpha=1,\dots,Q,
\end{equation*}
monic of degree $Q$, and the $NQ$ roots are all the moduli: the $Q$ roots of $P_a$ are the species-$a$ particles of Fig.~\ref{fig:picture}, and $\calM_Q\simeq\prod_a{\rm Sym}^Q\CC$ minus the locus of common roots. Self-duality is the linear Cauchy--Riemann equation in unnormalized variables, so multi-instantons are products of polynomials and do not interact classically. The trap $H$ is the moment map of the rotation $z\to e^{\ii t}z$, which has a single fixed point, all charge at the origin, and acts on each species with the weights $\{1,\dots,Q\}$ of $Q$ identical particles in the plane. The integral \eqref{eq:Zmod2d} is therefore of Duistermaat--Heckman type \cite{DH} and equals the Gaussian integral about the fixed point:
\begin{align}
 Z_Q&=\frac{1}{(Q!)^N\eps_1^{NQ}},\nonumber\\
 Z&=\sum_{Q\ge0}\frac{1}{(Q!)^N}\Big(\frac{u}{\eps_1^N}\Big)^{Q}=\hyp\Big(;1,\dots,1;\frac{u}{\eps_1^N}\Big),
 \label{eq:ZF}
\end{align}
where $\hyp$ is the hyper-Bessel function. For $N=2$, it is the Bessel function $I_0(2\sqrt u/\eps_1)$.

For a pedagogical derivation, see the longer version \cite{SDSYM}.
We note that the same partition function \eqref{eq:ZF} also appears as the vortex partition function of the $\CP$ model in an $\Omega$-background \cite{DGH,BTZ,FKNO} (see also \cite{Givental}), and the calculation is parallel.


This structure of $Z_Q$ offers the interpretation of the fractional constituents: $Z_Q$ is a gas of $N$ species, each with $Q$ free particles of fugacity $u^{1/N}$ trapped within the $\epsilon_1^{-1}$ box. The vacuum is the grand canonical sum over $Q$, which is not $N$ independent gases: with the fugacity for the fractional constituent $y=u^{1/N}/\eps_1$,
\begin{equation}
 Z=\sum_{Q\ge0}\frac{y^{NQ}}{(Q!)^N}=\int_0^{2\pi}\prod_{a=1}^{N-1}\frac{\dd\alpha_a}{2\pi}\,\exp\Big(y\sum_{a=1}^{N}e^{\ii\alpha_a}\Big),
 \label{eq:torus}
\end{equation}
with $\alpha_N\equiv-\sum_{a<N}\alpha_a$. 
Setting $u_0 = (C_N\,m\Lambda)^N > 0$ so that $u = u_0 e^{\ii\theta}$, we can write $y=y_\ell = \frac{u_0^{1/N}}{\eps_1} e^{\ii (\theta + 2 \pi \ell)/N}$, and there is the ambiguity to choose the branch $\ell = 0,\cdots, N-1$. The partition function itself (\ref{eq:torus}) does not depend on the choice of $\ell$ and is well-defined.

The phases enforce equal numbers of each species, which is integer topological charge. Equation \eqref{eq:torus} is quite parallel to the self-dual restriction of the mini-space sum of \cite{Unsal2020} for the $N$ types of monopole-instantons on $\RR^3\times S^1$, with the same constraint, under a suitable identification of parameters $\zeta V\leftrightarrow u^{1/N}/\eps_1$, where $\zeta$ is the monopole fugacity. 
This instanton-only sector leads to the pseudoconstituent gas $\sum_Q(\zeta V)^{NQ}/(Q!)^N$ of \cite{NU}; here, instead of the constituent gas, it arises on $\RR^2$, directly from the moduli space of integer instantons.

In the thermodynamic limit $\epsilon_1 \rightarrow 0$, which is $|y| \rightarrow \infty$, the saddle points $e^{\ii\alpha_a}=e^{2\pi\ii\ell/N}$ give
\begin{align}
 \lim_{\epsilon_1 \rightarrow 0} (-\epsilon_1 \log Z )&= \min_{\ell \in \mathbb{Z}_N} \left(-N u_0^{1/N}e^{\ii (\theta + 2 \pi \ell)/N} \right) 
 \label{eq:asym2d}
\end{align}
where $\min_{\ell \in \mathbb{Z}_N}$ refers to the real part.
This demonstrates that the multibranch structure emerges from the instanton sum.
There are $N$ branches, each branch exhibits the fractional $\theta$ dependence, and the phase transition occurs at $\theta = \pi$.

For example, for $N=2$, the partition function is given by $Z=I_0(2\sqrt u/\eps_1)$: every term of the series is an integer power of $u\propto e^{\ii\theta}$, yet $I_0(x)\simeq(e^x\pm\ii e^{-x})/\sqrt{2\pi x}$ grows with $\sqrt u\propto m\Lambda\,e^{\ii\theta/2}$. The two signs of the root are the branches $\theta/2$ and $(\theta+2\pi)/2$, and the vacuum is the one with the larger real part.

On this phase transition, we can give a picture from the Lee--Yang zeros.
At $\theta=0$, $Z$ is a sum of positive terms; for general $\theta$ the terms are complex; at $\theta=\pi$ the sum becomes an oscillatory function, $J_0(2\sqrt{|u|}/\eps_1)$, whose zeros accumulate on the real $\theta$ axis as the area grows. With $Y=\sqrt{|u|}/\eps_1$, proportional to the area, the zeros are at $\theta_k=\pi+2\ii\log(2Y/j_{0,k})$, with $j_{0,k}$ the zeros of $J_0$, on the line ${\rm Re}\,\theta=\pi$ with spacing $\simeq\pi/Y$. This is the first-order transition at $\theta=\pi$, seen through the Lee--Yang zeros of the partition function \cite{YangLee,LeeYang}.


To calculate the vacuum energy density, we assume locality for a large trap: 
\begin{equation}
 \log Z\simeq - \int\dd^2x\,\calE\big(u(x)\big),
\end{equation}
where $u(x)=u\,e^{-\eps_1|z|^2}$ and $\calE$ is the free energy density and $\calE(u(\infty))=0$.
Since $u\partial_u \calE(u(x))=-\frac{1}{2\eps_1|z|}\partial_{|z|} \calE(u(x))$, we can extract the free energy density near origin by
\begin{equation}
 \calE(u) = -\frac{1}{\pi}u\partial_u\lim_{\eps_1\to0}\eps_1\log Z.
\end{equation}

The vacuum energy density on branch $\ell$ is,
\begin{equation}
 \calE_\ell(\theta)=-\frac{ C_N\,m\Lambda}{\pi}\;e^{\ii(\theta+2\pi\ell)/N},\qquad \ell=0,\dots,N-1,
 \label{eq:E2d}
\end{equation}
which yields the chiral condensate $\langle\bpsi_+\psi_-\rangle_\ell=\partial_m\calE_\ell\propto\Lambda\,e^{\ii(\theta+2\pi\ell)/N}$.
This corresponds to the holomorphic part of the chiral condensate in the $\mathcal{N}=(2,2)$ $\CP$ model.
This also indicates a topological density $\langle q\rangle=u^{1/N}/N\pi$ of the self-dual vacuum.

The dilute gas keeps one position per instanton with weight $u$ and cuts the size integral at $\rho\sim\Lambda^{-1}$ by hand. It gives $\log Z/V\propto\Lambda^2(m/\Lambda)^N e^{\ii \theta}$ with a volume $V$, which is extensive but with a single branch. The full sum keeps all $N$ positions, each with weight $u^{1/N}$, and no cut-off. Expanded in powers of $u$, as a dilute expansion would do, each term of the exact $\log Z$ grows faster than the area, $\log I_0(2y)=y^2-y^4/4+\dots$, while the sum is $2y-\frac12\log(4\pi y)$. This is the infrared embarrassment \cite{ColemanBook,Witten79} in regulated form: the one-instanton term is $u(V/\pi)^N$, with the trap area $V=\pi/\eps_1$, and the expansion in $u$ is the wrong expansion.

\section{Four dimensions: self-dual $\mathcal N=1$ SYM, the ADHM gas and the Nekrasov--Shatashvili limit}
The four-dimensional theory is the Chalmers--Siegel self-dual Yang--Mills theory \cite{ChalmersSiegel} with one chiral gluino $\lambda_\alpha$ and a gluino mass, on $\RR^4=\CC^2\ni(z_1,z_2)$:
\begin{align}
 S&=\int\!\dd^4x\,\tr\Big[\ii B^{\dot\alpha\dot\beta}F_{\dot\alpha\dot\beta}+\bar\lambda_{\dot\alpha}D^{\alpha\dot\alpha}\lambda_\alpha+\frac m2\lambda^\alpha\lambda_\alpha\Big]\nonumber\\
 &\quad+\int\!\dd^4x\,T(x)\,q(x),\qquad q=\frac{1}{8\pi^2}\tr F\wedge F,
 \label{eq:S4d}
\end{align}
with $T(x)=8\pi^2/g^2-\ii\theta+\eps_1|z_1|^2+\eps_2|z_2|^2$. The multiplier $B$ sets the anti-self-dual part $F_{\dot\alpha\dot\beta}$ to zero, and $\bar\lambda$ enters linearly; $\lambda_\alpha$ carries the zero modes, $2Nk$ of them in charge $k$. 
As in two dimensions, $m$ is a source for $\tr\lambda\lambda$: the physical gluino mass is $\sqrt{m\,m_2}$, with $m_2$ the coefficient of $\bar\lambda\bar\lambda$, and the self-dual limit is $m_2\to0$.
Again, the trap potential $T(x) q(x)$ plays the role of the IR regulator as the system consists of instantons. 

Pauli--Villars counting of the $4Nk$ bosonic and $2Nk$ fermionic zero modes gives $\mu^{4Nk-Nk}$, which with $e^{-8\pi^2k/g^2+\ii k\theta}$ is $\Lambda^{3Nk}e^{\ii k\theta}$, and the mass contributes $m^{Nk}$:
\begin{equation}
 u=(C_N\,m\Lambda^3)^Ne^{\ii\theta},\qquad\Lambda^{3N}=\mu^{3N}e^{-8\pi^2/g^2(\mu)},
 \label{eq:u4d}
\end{equation}
with $[u]=4N$. So $u^{1/N}\propto m\Lambda^3e^{\ii\theta/N}$ is an energy density, and large volume means $|u|^{1/N}\gg\eps_{1,2}^2$.

Integrating $B$ imposes self-duality exactly, so the path integral lives on the moduli space $\calM_k$ of charge-$k$ instantons, of real dimension $4Nk$; integrating $\bar\lambda$ gives $\delta(D\lambda)$.
On non-zero modes, the Jacobian of the constraint, the ghosts and the gluino determinant cancel pointwise on $\calM_k$: with $L=d^-_A\oplus d^*_A$ the linearized constraint plus gauge fixing, on a self-dual background $LL^\dagger=(-D^2)\otimes1_4$ and $(\bar\sigma\!\cdot\!D)(\bar\sigma\!\cdot\!D)^\dagger=(-D^2)\otimes1_2$, so that
\begin{equation}
 \frac{\det(-D^2)}{|\det{}'L|}\,\big|\det{}'(\bar\sigma\!\cdot\!D)\big|=\det(-D^2)^{\,1-2+1}=1.
 \label{eq:121}
\end{equation}

The mass sees only the $2Nk$ zero modes of $\lambda$, on which the mass term is an antisymmetric form $\Omega$.
Since the adjoint representation is real, charge conjugation gives the zero-mode space a quaternionic structure in which $\Omega$ is the holomorphic symplectic form of the hyperk\"ahler $\calM_k$ \cite{Osborn,Maciocia,Dorey:2002ik}, and therefore, in orthonormal coordinates $\xi$ on the zero modes,
$\int\dd^{2Nk}\xi\,e^{-\frac m2\xi\Omega\xi}=m^{Nk}\,{\rm Pf}\,\Omega=m^{Nk}$ up to a constant absorbed to $u$.
Hence, we have
\begin{equation}
Z = \sum_{k\ge0}u^k\!Z_k = \sum_{k\ge0}u^k\!\int_{\calM_k}\!\dd{\rm vol}\;e^{-H},
 \label{eq:Zmod4d}
\end{equation}
with $H=\int\dd^4x\,(\eps_1|z_1|^2+\eps_2|z_2|^2)\,q(x)$.
See \cite{Dorey:2002ik} for the instanton measure.

\emph{The ADHM representation.} 
The moduli integration can be rewritten as the integration over the ADHM variables.
Osborn's formula for the second moments of $q$ \cite{Osborn} expresses $H$ in ADHM data
, which is the moment map of the rotations $z_a\to e^{\ii t_a}z_a$. Since the $L^2$ metric is also the hyperk\"ahler quotient of the flat ADHM metric \cite{Maciocia},  $Z_k$ is the equivariant volume of $\mathcal M_k$, i.e. Nekrasov's $k$-instanton coefficient \cite{Nekrasov,NekrasovOkounkov},  for the  $\mathcal N=2$ Seiberg--Witten theory \cite{Seiberg:1994rs}, in the limit $a\to0$ of the Coulomb parameter.
The only difference is the meaning of $\eps_{1,2}$: here they are couplings of the four-dimensional Lagrangian rather than parameters of an $\Omega$-background. The localization computation of \cite{Nekrasov,MNS,NS} therefore applies unchanged and gives a positive one-dimensional gas:
\begin{align}
 Z_k&=\frac{1}{k!}\Big(\frac{\eps}{2\pi\eps_1\eps_2}\Big)^{k}\!\int_{\RR^k}\!\dd^ky\;\frac{\prod_{i<j}K(y_i-y_j)}{\prod_{i=1}^{k}\big(\eps^2/4+y_i^2\big)^{N}},\nonumber\\
 K(d)&=\frac{d^2(d^2+\eps^2)}{(d^2+\eps_1^2)(d^2+\eps_2^2)},\qquad\eps=\eps_1+\eps_2.
 \label{eq:gas}
\end{align}

\emph{The NS-type thermodynamic limit.} 
We now have an analytic expression of each $Z_k$ in the instanton sum.
In principle, the partition function is calculable.

As in two dimensions, we define the energy density $\calE$ by assuming the locality relation $\log Z\simeq-\int d^4x\,\calE(u(x))$ with $u(x)=u\,e^{-\epsilon_1|z_1|^2-\epsilon_2|z_2|^2}$, and we can extract it as 
\begin{equation}
    \calE(u) = -\frac{1}{\pi^2}(u\partial_u)^2\lim_{\epsilon_1\to0}\lim_{\epsilon_2\to0}\epsilon_1\epsilon_2\log Z.
\end{equation}

In End Matter, we evaluate $\log Z$ in the Nekrasov–Shatashvili-type limit ($\epsilon_1, \epsilon_2 \rightarrow 0$ with $\epsilon_1 \gg \epsilon_2$), and we then obtain the vacuum energy as a function of $\theta$:
\begin{align}
 \calE(\theta)&= \min_{\ell \in \mathbb{Z}_N} \left[-\frac{|u|^{1/N}}{\pi^2}\,e^{\ii(\theta+2\pi\ell)/N}\right],
 \label{eq:E4d}
\end{align}
where $\min_{\ell\in\mathbb Z_N}$ means choosing the branch for which the real part of the expression is smallest.
Precisely speaking, it is straightforward to derive the $\sim u^{1/N}$ behavior only for $u > 0$. 
Continued to small $\theta$ and combined with the $2\pi$ periodicity, this already requires a switch of branches somewhere in $0<\theta<2\pi$.
However, the complete $\theta$ dependence, (\ref{eq:E4d}) for complex $u$, needs an assumption that the continuation of the $u>0$ saddle point is not overtaken by any other saddle point.
The connection of the saddle-point equation to the Baxter equation of the periodic Toda chain \cite{NS, KozlowskiTeschner} may provide a complementary route to establishing the complete $\theta$ dependence.
We revisit these points in the follow-up paper \cite{SDSYM}.

This vacuum energy is linear in $m\Lambda^3$. The resulting gluino condensate on the $\ell$-th branch of \eqref{eq:E4d}, $\langle\tr\lambda\lambda\rangle_\ell=2\,\partial_m\calE_\ell\propto\Lambda^3e^{\ii(\theta+2\pi\ell)/N}$, is consistent with the holomorphic part of the known condensate and the $N$ phases of $\mathcal N=1$ SYM \cite{NSVZ,ShifmanVainshtein88,DaviesHKM}.
As inferred from the energy-versus-entropy argument \cite{HU}, the self-dual vacuum forms a macroscopic condensate of the topological charge, with density $\langle q\rangle=u^{1/N}/N\pi^2$.



\section{Discussion}

\emph{Multibranch structure from instanton sum.}
Self-dual SYM with a chiral mass deformation can be regarded as a simplified version of self-dual Yang--Mills theory in which we can study the instanton sum without the complications of the full fluctuation determinant. 
We have demonstrated that this instanton sum produces fractional $\theta$ dependence and a multibranch structure, as stated in the Introduction.
The free energies of $N$ branches depend on $(\theta+2\pi\ell)/N$, with the dominant branch selected by the real part of the free energy with the transition at $\theta=\pi$. 
In these theories, the mismatch between integer instanton charge and the $N$-branched $\theta$ dependence, raised in 1979 \cite{Witten79}, is resolved by the sum over all instanton sectors, without fractional configurations.

We emphasize that the fractional dependence is a collective property of the instanton sum.
We do not assume fractional instantons, but the moduli space volume has the interpretation as fractional constituents, e.g., (\ref{eq:torus}).
As in Nekrasov's instanton counting \cite{Nekrasov,NekrasovOkounkov}, which reproduces the Seiberg–Witten solution \cite{Seiberg:1994rs}, integrating over the full moduli space allows us to go beyond the dilute-gas description.

\emph{Semiclassics.}
In the usual instanton calculation, the integral over the size modulus is infrared divergent: the running coupling $g^2(1/\rho)$ grows with the instanton size $\rho$, and the weak-coupling expansion becomes unreliable. 
In the present theory, the one-loop-exact coupling $g_\sdd^2(1/\rho)$ also grows at large $\rho$. 
Here, however, each topological sector is given by a coupling-independent moduli-space integral, with the coupling entering through the dynamical scale $\Lambda$. 
The same decomposition holds in bosonic self-dual Yang--Mills
theory~\cite{ChalmersSiegel}, where each sector is weighted by
't~Hooft's one-loop determinant $\det(-D^2_{\rm adj})^{-1}$~\cite{tHooft76}
on the moduli space: still exact decomposition, but no longer a volume.
In the present theory
this reduction allows us to evaluate the sum beyond the weak-coupling dilute gas.

In suitably compactified theories, the multibranch structure can be derived in a weakly coupled regime using a dilute gas of fractional instantons \cite{Unsal2008,ShifmanUnsal2008,Unsal2009,DunneUnsalReview,TanizakiUnsal2022}.
Unlike in compactified semiclassics, where fractional instantons generate the fractional $\theta$ dependence, in our calculation it emerges from the sum over integer-charge instantons. Exploring the relation between these two descriptions will be investigated in the future work.

\emph{Further directions.}
For the two-dimensional $\mathbb{C}P^{N-1}$ model, the integral representation of the instanton sum (\ref{eq:torus}) exhibits an affine Toda structure closely resembling that of the Hori--Vafa mirror superpotential \cite{HV}. This resemblance suggests a connection between the instanton sum and the mirror description that would be interesting to explore further.

Another question is how the results change when the simplifications of the present theory are relaxed \cite{DGNU}. Restoring the full fluctuation determinant of self-dual Yang--Mills theory would test the persistence of the multibranch structure, while including anti-instantons would allow us to investigate its relation to the vacuum structure of the full theory. The success of compactified semiclassics also motivates asking whether this approach can account for confinement and the mass gap. A longer account will appear in \cite{SDSYM}.

\begin{acknowledgments}
We thank Yuya Tanizaki and Philip Argyres for discussions.
The authors used the assistance of Claude Opus 5.5, Claude Fable 5.1, ChatGPT 6.1 Sol, and ChatGPT 6 Astra for exploratory calculations, cross-checks, and proofreading.
M.~\"U. is supported by the U.S. Department 
of Energy, Office of Science, Office of Nuclear Physics under 
Award Number DE-FG02-03ER41260, and the Simons Foundation Grant ``Simons Collaboration on Confinement 
and QCD Strings''. 
\end{acknowledgments}

\bibliography{fractional}

\clearpage
\section*{End Matter}
\appendix

\section{Evaluation of instanton sum}\label{app:leading}

We consider the large-volume limit $\epsilon_1,\epsilon_2\to0$ of the instanton sum \eqref{eq:Zmod4d}. 
To simplify the computation, we take it in the hierarchy of the Nekrasov--Shatashvili limit \cite{NS}, $\epsilon_2\to0$ first at fixed $\epsilon_1$, and then $\epsilon_1\to0$.
We first show that, for $u>0$, the free energy can be evaluated as,
\begin{equation}
\lim_{\epsilon_1\to0}\ \lim_{\epsilon_2\to0}\ \epsilon_1\epsilon_2\log Z=N^2u^{1/N}.
\end{equation}
Then, under an assumption that no saddle point other than the continuation of the $u>0$ one contributes for complex $u$, we can determine the complete $\theta$ dependence of the vacuum energy.

Let us begin by rewriting (\ref{eq:gas}).
We split the two-body factor as
\begin{equation}
K(d)=\frac{d^2}{d^2+\epsilon_2^2}\,e^{U(d)},\qquad U(d)=\log\frac{d^2+\epsilon^2}{d^2+\epsilon_1^2}.
\end{equation}
The first factor gives a Cauchy determinant,
\begin{equation}\label{eq:cauchy}
\prod_{i<j}\frac{(y_i-y_j)^2}{(y_i-y_j)^2+\epsilon_2^2}=\det_{i,j}\frac{\epsilon_2}{\epsilon_2+i(y_i-y_j)} .
\end{equation}
The Fourier transform of $U$, $\tilde U(q)=\int\mathrm{d} y\,e^{-iqy}U(y)$, is
\begin{equation}\label{eq:Utilde}
\tilde U(q)=\frac{2\pi}{\lvert q\rvert}e^{-\epsilon_1\lvert q\rvert}\big(1-e^{-\epsilon_2\lvert q\rvert}\big)>0 .
\end{equation}
So, with a Gaussian field $\langle\varphi(y)\varphi(y')\rangle_\varphi=U(y-y')$,
\begin{equation}
\prod_{i<j}e^{U(y_i-y_j)}=\Big\langle e^{\sum_i(\varphi(y_i)-\frac12U(0))}\Big\rangle_\varphi .
\end{equation}
The sum over $k$ in \eqref{eq:Zmod4d} then becomes
\begin{align}\label{eq:fredholm}
Z&=\sum_{k\ge0}\frac1{k!}\Big\langle\int\mathrm{d}^ky\prod_iA(y_i)\,\det_{i,j}\hat K(y_i,y_j)\Big\rangle_\varphi\nonumber\\
&=\langle\det(1+A\hat K)\rangle_\varphi,\nonumber\\
\hat K(y,y')&=\frac{1}{2\pi(\epsilon_2+i(y-y'))},\ \
A(y)=\frac{u\,e^{\varphi(y)}}{(\epsilon^2/4+y^2)^{N}},
\end{align}
with $\det(1+A\hat K)$ defined by the series, where $\det$ denotes the functional determinant.

\textit{Effective action of $\varphi$.}
By \eqref{eq:fredholm}, $Z$ is an integral over $\varphi$ with the effective action
\begin{align}
&Z=\int D\varphi\,e^{-S[\varphi]},\nonumber\\
&S[\varphi]=\frac12\int\frac{\mathrm{d} q}{2\pi}\frac{\lvert \tilde\varphi(q)\rvert^2}{\tilde U(q)}-\log\det(1+A\hat K).
\end{align}
We compute $S$ as $\epsilon_2\to0$.
By \eqref{eq:Utilde}, $\epsilon_2/2\tilde U(q)=e^{\epsilon_1\lvert q\rvert}\big(1+O(\epsilon_2q)\big)/4\pi$.
For the determinant, expanding $\operatorname{tr}\log(1+A\hat K)$ in powers of $A\hat K$ and resumming gives\footnote{
The kernel $\hat{K}$ is diagonal in plane waves: $\hat K e^{-ipy}=e^{-\epsilon_2p}\theta(p)\,e^{-ipy}$, where $\theta(p)$ is the step function. Taking the trace of $(A\hat K)^m$ in plane waves, the $p$ integral gives $1/2\pi m\epsilon_2$ as $\epsilon_2\to0$, so that $\epsilon_2\operatorname{tr}(A\hat K)^m\to\frac{1}{2\pi m}\int\mathrm{d} y\,A^m$; and $\sum_m\frac{(-1)^{m+1}}{m^2}A^m=-\mathrm{Li}_2(-A)$.}
\begin{equation*}
\lim_{\epsilon_2\to0}\epsilon_2\log\det(1+A\hat K)=\frac{1}{2\pi}\int\mathrm{d} y\,\big[-\mathrm{Li}_2(-A)\big].
\end{equation*}
Together, $S[\varphi] \simeq -J[\varphi]/\epsilon_2$ with
\begin{align}\label{eq:J}
J[\varphi]=&-\frac1{4\pi}\int\frac{\mathrm{d} q}{2\pi}\,e^{\epsilon_1\lvert q\rvert}\lvert \tilde\varphi(q)\rvert^2\nonumber\\
&-\frac1{2\pi}\int\mathrm{d} y\,\mathrm{Li}_2\Big(-\frac{u\,e^{\varphi}}{(y^2+\epsilon_1^2/4)^N}\Big).
\end{align}

This $J$ is the functional of the Nekrasov--Shatashvili limit \cite{NS,MY}.
Thus the $\varphi$-integral is evaluated at the saddle point $\delta J/\delta\varphi=0$, where $\epsilon_2\log Z\to J$.

\textit{Saddle point.}
We change variables from $\varphi$ to
\begin{equation*}
\rho\equiv\frac{\epsilon_1}{2\pi}e^{\epsilon_1 |\partial|}\varphi,
\end{equation*}
which stays finite as $\epsilon_1\to0$.\footnote{\label{footnote:density}
Indeed, $\rho$ can be regarded as the density.
In \eqref{eq:fredholm} each $y_i$ carries $e^{\varphi(y_i)}$, so $\epsilon_1\epsilon_2\langle\sum_i\delta(y-y_i)\rangle=\epsilon_1\epsilon_2\langle\delta\log\det(1+A\hat K)/\delta\varphi(y)\rangle$, which tends to $\rho$ at the saddle point. Since the number of instantons is expected to grow as the volume of the trap, $\propto1/\epsilon_1\epsilon_2$, we look for $\rho$ that stays finite.}

With $(|\partial|f)(y)=\int\frac{\mathrm{d} q}{2\pi}e^{iqy}|q|\tilde f(q)$, the saddle-point equation $\delta J/\delta\varphi=0$ reads
\begin{equation}
\frac{2\pi}{\epsilon_1}\rho=\log(1+A). \label{eq:saddle}
\end{equation}
Equivalently, we have
\begin{equation}
\log\frac{u}{(y^2+\epsilon_1^2/4)^N}-\frac{2\pi}{\epsilon_1}\big(1-e^{-\epsilon_1|\partial|}\big)\rho=\log\big(1-e^{-2\pi\rho/\epsilon_1}\big). \label{eq:saddle-log}
\end{equation}
For $u>0$, (\ref{eq:saddle}) implies $\rho>0$ because $A>0$.
As $\epsilon_1\to0$, $\rho$ is either of order one or of order $\epsilon_1$ at each $y$; below, $\rho$ denotes its part of order one, and $(1-e^{-\epsilon_1|q|})/\epsilon_1\to|q|$.
Where $\rho$ is of order one, the right side of \eqref{eq:saddle-log} vanishes, and
\begin{equation}\label{eq:airfoil-eq}
\log u-N\log y^2=2\pi|\partial|\rho\qquad(\rho>0).
\end{equation}
When $\rho$ is of order $\epsilon_1$, \eqref{eq:saddle-log} has a solution only if its left side is non-positive: $\log u-N\log y^2\le2\pi|\partial|\rho$.
These are the conditions for $\rho\ge0$ to maximize the following functional, which is the limit of $\epsilon_1 J$ as $\epsilon_1 \to 0$:\footnote{
Where $\rho$ is positive and of order one, use $-\mathrm{Li}_2(-A)=\frac12\log^2A+\frac{\pi^2}6+\mathrm{Li}_2(-1/A)$ with the principal branch of $\log A$ ($|\mathrm{Im}\log A|<\pi$), and the terms of order $1/\epsilon_1$ cancel. Where $\rho$ is negative and of order one, $A\to0$ and $\epsilon_1J\to-\infty$.
}
\begin{equation}
\int\mathrm{d}y\,\rho\,(\log u-N\log y^2)-\pi\int\mathrm{d}y\,\rho|\partial|\rho.  \label{eq:functional}
\end{equation}
This functional  is concave in $\rho$ since $|q| \geq 0$, so that the maximum is unique.
$\epsilon_1\epsilon_2\log Z$ can be evaluated by its maximum.

\textit{Solution for $u>0$.}
We introduce dimensionless variables by $y=u^{1/2N}t$ and $\rho=Nu^{1/2N}g(t)$; then \eqref{eq:airfoil-eq} reads $\pi\lvert \partial\rvert g=-\log\lvert t\rvert$.
Its solution with $g\ge0$, and $g$ and $g'$ vanishing at the edges, is
\begin{align}\label{eq:airfoil}
\rho(y)&=Nu^{1/2N}g\big(y/u^{1/2N}\big),\nonumber\\
g(t)&=\frac1\pi\Big[\sqrt{4-t^2}-\lvert t\rvert\arccos\frac{\lvert t\rvert}2\Big]
\end{align}
for $\lvert t\rvert\le2$, and $g=0$ otherwise.
This is similar in form to the Logan--Shepp--Vershik--Kerov limit shape \cite{LS,VK}.
One can check that \eqref{eq:airfoil} satisfies both conditions.
At the maximum, \eqref{eq:functional} equals $\pi\int\mathrm{d}y\,\rho\lvert\partial\rvert\rho$ by \eqref{eq:airfoil-eq}, and
\begin{equation*}
\epsilon_1\epsilon_2\log Z\to\pi\int\mathrm{d} y\,\rho\lvert \partial\rvert\rho=N^2u^{1/N}.
\end{equation*}

\textit{Towards complex $u$.}
The fractional power $u^{1/N}$ already suggests the multibranch structure.
For complex $u$, the saddle point of $J$ is hard to find.
However, in \eqref{eq:functional}, $\log u$ couples only to $\int\mathrm{d}y\,\rho$.
We thus reduce the problem to the single variable $M=\int\mathrm{d}y\,\rho$.
We fix $\int\mathrm{d}y\,\rho=M$ with a Lagrange multiplier $\lambda$, whose original contour is imaginary.
Here $\lambda$ plays the role of $\log u$, which multiplies the instanton number in \eqref{eq:Zmod4d} with $Z_k>0$, so, at real $M$ on the original contour, we expect the dominant saddle point to be at real $\lambda$.
There, the saddle-point equation for $\rho$ is \eqref{eq:airfoil-eq} with $\lambda$ in place of $\log u$, so that $\rho$ is \eqref{eq:airfoil} with $u=e^\lambda$, and then we take the saddle point in $\lambda$, $Ne^{\lambda/N}=M$:
\begin{align}\label{eq:ZkM}
Z&\simeq\int\mathrm{d}M\int\mathrm{d}\lambda\int D\rho\,\exp\frac1{\epsilon_1\epsilon_2}\Big[M(\log u-\lambda)\nonumber\\
&\hspace{3em}+\int\mathrm{d}y\,\rho\,(\lambda-N\log y^2)-\pi\int\mathrm{d}y\,\rho\lvert\partial\rvert\rho\Big]\nonumber\\
&\xrightarrow{\ \rho\ }\int\mathrm{d}M\int\mathrm{d}\lambda\,\exp\frac1{\epsilon_1\epsilon_2}\Big[M(\log u-\lambda)+N^2e^{\lambda/N}\Big]\nonumber\\
&\xrightarrow{\ \lambda\ }\int\mathrm{d}M\,\exp\frac1{\epsilon_1\epsilon_2}\Big[M\log u+NM\Big(1-\log\frac MN\Big)\Big].
\end{align}
Note that $Z$ does not depend on the branch of $\log u$, which appears explicitly in \eqref{eq:functional} and \eqref{eq:ZkM}.
The apparent branch dependence is due to our choice of the principal branch, $|\theta|<\pi$, for $u = |u|e^{\ii \theta}$ in deriving \eqref{eq:functional}.
The saddle point of \eqref{eq:ZkM} in $M$ is then at $M=Nu^{1/N}$, which is complex for $\theta\neq0$.
Continuing the real-$M$ result to this point, we assume that it is not overtaken by any other saddle point for $|\theta|<\pi$, which gives $\epsilon_1\epsilon_2\log Z\to N^2| u|^{1/N}e^{i\theta/N}$.
As $Z$ does not depend on the branch, this holds for any $\theta$ once we return to the principal branch by $\log u\to\log u+2\pi i\ell$, so that
\begin{equation}
\epsilon_1\epsilon_2\log Z\to\max_{\ell\in\mathbb Z_N}N^2\lvert u\rvert^{1/N}e^{i(\theta+2\pi\ell)/N},
\end{equation}
where the maximum refers to the real part.
Its real part has a cusp at $\theta=\pi$, where the maximum switches between two of the $N$ branches.
This is the multibranch structure.
It holds unless another saddle point overtakes it for some $|\theta|<\pi$, which we leave to future work.

\end{document}